\documentstyle[aps,multicol,epsf,amsbsy,cite]{revtex}

\newcommand{\kB}{k_{\mathrm{B}}}

\begin{document} 
\title{Magnetization Switching in Nanowires: Monte Carlo Study with
  Fast Fourier Transformation for Dipolar Fields}

\author{D.~Hinzke and U.~Nowak}
\address{Theoretische Tieftemperaturphysik,
  Gerhard-Mercator-Universit\"{a}t Duisburg, 47048 Duisburg/ Germany\\ 
  e-mail: denise@thp.uni-duisburg.de, uli@thp.uni-duisburg.de}

\date{February 9, 2000} \maketitle

\begin{abstract}
  For the investigations of thermally activated magnetization reversal
  in systems of classical magnetic moments numerical methods are
  desirable.  We present numerical studies which base on time
  quantified Monte Carlo methods where the long-range dipole-dipole
  interaction is calculated with the aid of fast Fourier
  transformation.  As an example, we study models for ferromagnetic
  nanowires comparing our numerical results for the characteristic
  time of the reversal process also with numerical data from Langevin
  dynamics simulations where the fast Fourier transformation method is
  well established.  Depending on the system geometry different
  reversal mechanism occur like coherent rotation, nucleation, and
  curling.
\end{abstract}

Pacs: 75.10.Hk, 75.40.Mg, 75.40.Gb


\begin{multicols}{2}

\section{Introduction}

Small magnetic particles in the nanometer regime are interesting for
fundamental research as well as for applications in magnetic devices.
With decreasing size thermal activation becomes more and more important
for the stability of nanoparticles and is hence investigated
experimentally as well as theoretically.  Wernsdorfer et al.\ studied
magnetization reversal in nanowires
\cite{wernsdorferPRL96} as well as in nanoparticles
\cite{wernsdorferPRL97} experimentally. They found
that for very small particles the magnetic moments rotate coherently
as in the Stoner-Wohlfarth model \cite{wernsdorferPRL97} while for
larger system sizes more complicated nucleation processes occur.

Numerical studies of the thermal activation in magnetic systems base
either on Langevin dynamics
\cite{lyberatosJAP93,nakataniJMMM97,garciaPRB98,zhangJAP99} or on
Monte Carlo methods. With the second method, mainly nucleation
processes in Ising systems have been studied
\cite{rikvoldBOOK98,acharyyaEPJ98}, but also vector spin models have
been used to investigate the switching behavior in systems with
continuous degrees of freedom
\cite{gonzalesJAP97,hinzkePRB98,nowakJAP99,nowakPRL00,hinzkePRB00}.
In these numerical studies the dipole-dipole interaction is often
neglected or at least approximated due to its large computational
effort.

In this work, we will use Monte Carlo methods in order to investigate
the thermally activated magnetization behavior of a systems of
magnetic moments including the dipole-dipole interactions.  Since the
calculation of these long-range interaction is extremely time
consuming when performed straightforward it is necessary to develop
efficient numerical methods for this task. We will demonstrate that
the implementation of fast Fourier transformation (FFT) methods ---
which are already established in the context of micromagnetic
simulations based on the Landau-Lifshitz equation of motion
\cite{yuanIEEE92,berkovPSS93} --- is possible also in a Monte Carlo
algorithm where a single-spin-flip method is used.
 
Our approach is applied to the important problems of thermally
activated magnetization reversal in a model for nanowires.  Depending
on the system geometry, material parameters, and the magnetic field
different reversal mechanisms occur.  Very small wires reverse by a
coherent rotation mode while for sufficiently long wires the reversal
is dominated by nucleation \cite{braunBOOK00,hinzkePRB00}. With
increasing system width an additional crossover sets in to a reversal
by a curling-like mode.  Our numerical results based on Monte Carlo
simulations are supplemented by Langevin dynamics simulations for
comparison.

\section{Model for a nanowire}

We consider a classical Heisenberg Hamiltonian for localized spins on
a lattice,
\begin{eqnarray}
  \label{e:ham}
 {\cal H} &=& - J \sum_{\langle ij \rangle} {\mathbf S}_i \cdot {\mathbf
    S}_j - \mu_s{\mathbf B}\cdot \sum_i {\mathbf S}_i \nonumber \\ &-&
  w \sum_{i<j} \frac{3({\mathbf S}_i \cdot {\mathbf e}_{ij})({\mathbf
      e}_{ij} \cdot {\mathbf S}_j) - {\mathbf S}_i \cdot {\mathbf
      S}_j}{r^3_{ij}},
\end{eqnarray}
where the ${\mathbf S}_i = {\boldsymbol \mu}_i/\mu_s$ are three
dimensional magnetic moments of unit length. The first sum represents
the ferromagnetic exchange of the moments where $J$ is the coupling
constant, the second sum is the coupling of the magnetic moments to an
external magnetic field $B$, and the last sum represents the dipolar
interaction. $w = \mu_0 \mu_s^2 /(4 \pi a^3)$ describes the strength
of the dipole-dipole interaction with $\mu_0 = 4 \pi \cdot 10^{-7}
{\mathrm Vs} ({\mathrm Am})^{-1}$. We consider a cubic lattice with lattice
constant $a$. The ${\mathbf e}_{ij}$ are unit vectors pointing
from lattice site $i$ to $j$.

We model a finite, cylindric system with diameter $D$ and length $L$
(in number of lattice sites). All our simulations start with a
configuration where all spins point up (into $z$ direction).  The
external magnetic field is antiparallel to the magnetic moments, hence
favoring a reversal process. As in earlier publications
\cite{hinzkePRB98,nowakJAP99,nowakPRL00,hinzkePRB00} we measure the
characteristic time of the reversal process, i.\ e.\ the time when the
$z$ component of the magnetization of the system changes its sign
averaged over either 100 (Langevin dynamics) or 1000 (Monte Carlo)
simulation runs.

\section{Numerical Methods}  

\begin{figure}
  \begin{center}
    ~ \\[5mm]
    \epsfxsize=8cm
    \epsffile{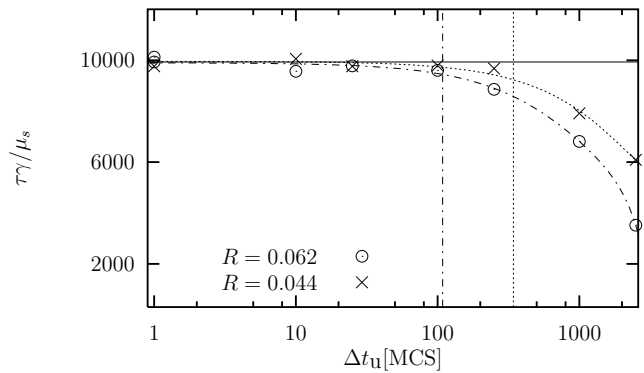}
  \end{center}
  \caption{Reduced characteristic time $\tau \gamma /\mu_{\mathrm s}$
    vs.\ update interval $\Delta t_{\mathrm u}$ of the dipolar fields
    for different trial step widths $R$. The solid line represents the
    ``correct'' value when the dipolar field is calculated after every
    spin update. The vertical lines represent the number of MCS which
    one spin needs at least to reverse.  Dashed lines connecting the
    points are guides to the eye.  $L = 8$, $\alpha = 1$, $\kB T =
    0.007J$, $\mu_s B_z = 0.15J$, and $w = 0.032 J$.}
 \label{f:tau_update}
\end{figure}

In the following we will mainly use a Monte Carlo method with a time
step quantification as derived in \cite{nowakPRL00} and later further
discussed in \cite{hinzkePRB00}. The method is based on a
single-spin-flip algorithm.  The trial step of the algorithm is a
random movement of the magnetic moment within a certain maximum angle.
In order to achieve this efficiently we construct a random vector with
constant probability distribution within a sphere of radius $R$ and
add this vector to the initial moment. Subsequently the resulting
vector is normalized. The trial step width $R$ is connected to a time
scale $\Delta t$ according to the relation
\begin{equation}
  R^2 = \frac{20 \kB T \alpha \gamma}{(1+\alpha^2) \mu_s} \Delta t.
\end{equation}
Using this relation one Monte Carlo step per spin (MCS) 
corresponds to the time interval $\Delta t$ of the corresponding
Landau-Lifshitz-Gilbert equation in the high damping limit
\cite{nowakPRL00}. Throughout the paper we set $\Delta t =
\frac{1}{40}\mu_s /\gamma$ and adjust $R$, except of the simulations
for Fig.~\ref{f:tau_update} where we
have to fix $R$ and calculate $\Delta t$.

In order to compute the dipole-dipole interaction efficiently 
we use FFT methods \cite{recipesBOOK90} for the calculation of the
long-range interactions \cite{yuanIEEE92,berkovPSS93}.  This method
uses the discrete convolution theorem for the calculation of the
dipolar field
\begin{equation}
{\mathbf H}_i = \sum_{j \neq i} \frac{3{\mathbf e}_{ij}({\mathbf
    e}_{ij} \cdot {\mathbf S}_j) - {\mathbf S}_j}{r^3_{ij}}.
\label{e:h_dip}  
\end{equation}
This dipolar field can be rewritten in the form
\begin{equation}
  H_i^{\eta} = \sum_{\theta, j} W_{ij}^{\eta \theta} S_j^{\theta}
\end{equation}
where the Greek symbols $\eta,\theta$ denote the Cartesian components
$x,y,z$ and the Latin symbols $i,j$ denote the lattice sites.
$(W^{\eta \theta})_{ij}$ are interaction matrices which only depend on
the lattice. Since the lattice is translational invariant one can apply
the discrete convolution theorem in order to calculate the Fourier
transform of the dipolar field as
\begin{equation}
  H_k^{\eta} = \sum_{\theta} W_k^{\eta \theta} S_k^{\eta}.
  \label{e:prod_k}
\end{equation}

Thus one first computes the interaction matrices $(W^{\eta
  \theta})_{ij}$ and its Fourier transform $(W^{\eta \theta})_k$. This
task has to be performed only once before the simulation starts since
the interaction matrices depend only on the lattice structure and,
hence, remain constant during the simulation.  For each given spin
configuration the dipolar field can then be calculated by first
performing the Fourier transform of the ${\mathbf S}_i$, second
calculating the product above following Eq.\ \ref{e:prod_k}, and third
transforming the fields ${\mathbf H}_k$ back into real space resulting
in the dipolar fields ${\mathbf H}_i$.  Using FFT techniques the
algorithm needs only of the order of $N \log N$ calculations
\cite{recipesBOOK90} instead of $N^2$ calculations which one would need
for the straightforward calculation of the double sum in Eq.\ 
\ref{e:ham}.

In order to apply the convolution theorem the system has
to be periodically and the range of the interaction has to be
identical to the system size. Since we are interested in finite systems
with open boundary conditions we use the zero padding method
\cite{recipesBOOK90}. This method is based on a duplication of the system
size where the original system is wrapped up with zero spins.

The FFT methods is well established as containing no further
approximation in the context of micromagnetism \cite{hubertBOOK98}.
There is however an additional problem concerning the implementation
of this method to Monte Carlo algorithms since here the spins are not
updated in parallel. Principally, in a Monte Carlo simulation the
dipolar field has to be calculated after each single spin flip since
the dipolar field at any site of the lattice depends on all other
magnetic moments. Thus if one magnetic moment moves, the value of the
whole dipolar field changes. Hence, in a usual Monte Carlo algorithm
one would store the dipolar fields in an array which is updated after
every accepted spin flip \cite{huchtJMMM95}. Here, only the changes of
the dipolar field due to the single spin update are calculated ($N$
operations) resulting in a need of computation time which scales with
$N^2$ per MCS. 

\narrowtext

Nevertheless, in the following we will show that alternatively one can
also recalculate the whole dipolar field at once after a certain
number of MCS taking advantage from the FFT method. This can be a good
approximation as long as the changes of the system configuration
during this update interval are small enough.

In order to investigate the validity of this idea systematically, we
consider thermally activated magnetization reversal in a spin chain,
in other words, in a cylinder with diameter $D=1$.  First, we
calculate the ``correct'' value of the reduced characteristic time
following from a Monte Carlo simulation where the dipolar field is
calculated after each accepted spin flip.  This value is shown as
solid line in Fig.~\ref{f:tau_update}.  The data points represent the
dependence of the reduced characteristic time on the length of the
update interval $\Delta t_u$ after which the dipolar field is
calculated. We used two different trial step widths $R$ for
comparison.

As is demonstrated our method converges already for update intervals
clearly longer than 1MCS, depending on the trial step width $R$. The
dependence on $R$ can be understood as follows.  The vertical lines
represent the minimal number of MCS which one spin needs to reverse.
This can be estimated from the mean step width of a magnetic moment
\cite{nowakPRL00} assuming that each trial step is accepted following
a strong external field. The mean step width within our Monte Carlo
procedure is $R / \sqrt{5}$ and thus the minimal number of MCS for a
spin reversal is $\pi \sqrt{5}/R$.  We conclude from this
considerations that it is a good approximation to update the whole
dipolar field after a certain interval of the Monte Carlo procedure
which has to be smaller than the minimum number of MCS needed for a
spin reversal.  Note that it follows, that this method should never
work for an Ising system.

\narrowtext
\begin{figure}[h]
  \begin{center}
    \epsfxsize=8cm
    \epsffile{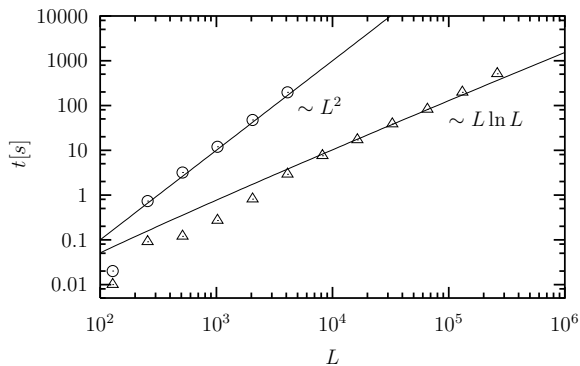}
  \end{center}
  \caption{Computation time $t$ vs.\ system size $L$ of a spin chain. The 
    data are from Monte Carlo simulations where the dipolar field is
    calculated either directly as explained in the text ($\circ$) or
    with FFT methods ($\triangle$).}
 \label{f:time_L}
\end{figure}

Now we will turn to an investigation of the efficiency and capability
characteristics of the method.  A comparison of the computation time
needed when the dipolar field is either calculated straightforward by
updating the field after each accepted spin flip or with the FFT
method after each MCS is shown in Fig.~\ref{f:time_L}.  Here, the
computation time is the CPU time needed on an IBM RS6000/590
workstation for 100MCS. Using the FFT method the computation time is
proportional to $L \ln{L}$ while for the usual method the computer
time scales like $L^2$. For the chain of length $2^{18} = 262144$ the
gain of efficiency of the FFT method is roughly a factor 5000 (note
that the FFT algorithm is most efficient for the system sizes which
are products of small prime numbers) which is a rather impressing
result.

As a test of the FFT method as well as the time quantification of the
Monte Carlo algorithm we use for comparison also another numerical
method namely Langevin dynamics simulations. Here we solve numerically
the Landau-Lifshitz-Gilbert equation of motion with Langevin dynamics
using the Heun method \cite{garciaPRB98}. This equation has the form
\begin{equation}
  \frac{(1+\alpha^2)\mu_s}{\gamma} \frac{\partial {\mathbf
  S}_i}{\partial t} = -{\mathbf S}_i \times \Big({\mathbf H}_i(t) +
  \alpha \big ({\mathbf S}_i \times {\mathbf H}_i(t)\big) \Big)
\end{equation}
with the internal field $ {\mathbf H}_i(t) = {\boldsymbol \zeta}_i -
\frac{\partial {\cal H}}{\partial {\mathbf S}_i}$, the gyromagnetic ratio
$\gamma = 1.76 \times 10^{11} ({\mathrm Ts})^{-1}$ and the
dimensionless damping constant $\alpha$.  The noise ${\boldsymbol
  \zeta}_i$ represents thermal fluctuations, with $\langle
{\boldsymbol \zeta}_i(t) \rangle = 0$ and $ \langle \zeta_i^{\eta}(t)
\zeta_j^{\theta}(t') \rangle = 2 \delta_{ij} \delta_{\eta \theta}
\delta(t-t') \alpha \kB T \mu_{s}/ \gamma$ where $i, j$ denote once
again lattice sites and $\eta, \theta$ Cartesian components.  The
implementation of the FFT method is straightforward for a Langevin
dynamics simulation since here a parallel update of spin
configurations is appropriate.
  
\narrowtext
\begin{figure}[h]
  \begin{center}
    \epsfxsize=8cm
    \epsffile{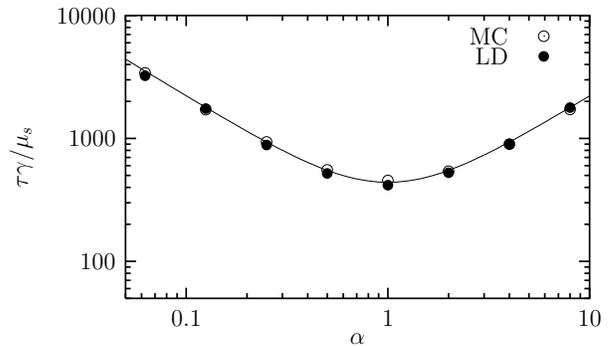}
  \end{center}
  \caption{Reduced characteristic time $\tau \gamma/\mu_{\mathrm s}$
    vs.~damping constant $\alpha$. The data are from Monte Carlo and
    Langevin dynamics simulations for a spin chain of length $L = 64$.
    The line resembles the $(1+ \alpha^2)/\alpha$ behavior. $\mu_s B_z
    = 0.15J$, $w = 0.032J$ and $\kB T = 0.025J$.}
  \label{f:tau_a_dz_0.0.ps}
\end{figure}

In order to compare our different methods, in
Fig.~\ref{f:tau_a_dz_0.0.ps} the $\alpha$ dependence of the
characteristic time for the reversal of a spin chain is shown. Monte
Carlo data are compared with those from Langevin dynamics simulations.
Interestingly, for the whole range of $\alpha$ values the Monte Carlo
and Langevin dynamics data coincide. This is in contrast to earlier
tests \cite{nowakPRL00,hinzkePRB00} where this agreement was achieved
only in the high damping limit. Seemingly, there exist certain systems
which show only a simple $\alpha$ dependence of the form $\tau \sim
(1+\alpha^2)/\alpha$. This $\alpha$ dependence usually describes the
high damping limit which is rendered by the time quantified Monte
Carlo algorithm.  In our model this $\alpha$ dependence is obviously
valid in the whole range of damping constants which we considered,
leading to the remarkable agreement. In general, this is not
necessarily the case since for other models different low and high
damping limits appear (see e.\ g.\ \cite{coffeyPRL98,CoffeyPRB98}),
depending on the systems properties. In the following we set $\alpha =
1$.

\section{Nucleation and Curling} 

As application of the methods described in the previous section we
start with a simple chain of classical magnetic moments of length $L =
128$ as extreme case of a cylindrical system with $D=1$. Thermally
activated magnetization reversal in a spin chain was treated
analytically by Braun \cite{braunPRL93,braunPRB94} who proposed a
soliton-antisoliton nucleation process as reversal mechanism.  For a
finite system with open boundary conditions the nucleation will
originate at the sample ends leading to an energy barrier
\cite{braunJAP99}
\begin{equation}
  \Delta E = 2\sqrt{2d}(\tanh{r}-hr)
  \label{e:e_nu}
\end{equation}
with $h = \mu_s B_z / (2 d)$ and $r = {\mathrm arcosh}{\sqrt{1/h}}$.
Here, the dipole-dipole interaction is approximated by a uniaxial
anisotropy of the form $-d \sum_i (S_i^z)^2$ with $w \approx d/\pi$
\cite{braunJAP99}. This estimate follows from a comparison of the
stray field energy of an elongated ellipsoid \cite{hubertBOOK98} with
the energy of a chain with uniaxial anisotropy.  In \cite{hinzkePRB00}
the occurrence of soliton-antisoliton nucleation was confirmed and
Eq.\ref{e:e_nu} was verified as well as the asymptotic form of the
corresponding characteristic time
\begin{equation}
 \tau = \tau_0 \exp{\frac{\Delta E}{\kB T}},
\end{equation}
where $\tau_0$ is a prefactor depending on the system parameters, the
applied magnetic field, and the damping constant. This prefactor was
also derived asymptotically for a periodic chain for the case of
nucleation \cite{braunPRB94} as well as for coherent rotation
\cite{braunJAP94}.  The latter is energetically favorable for low
enough system size. However, for a finite chain the prefactor is still
unknown.

\narrowtext
\begin{figure}
  \begin{center}
    \epsfxsize=8cm
    \epsffile{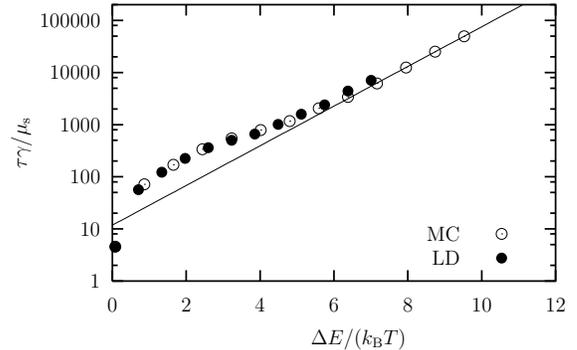}
       \end{center}
   \caption{Reduced characteristic time $\tau \gamma/\mu_{\mathrm s}$
     vs.~inverse temperature $\Delta E/ k_{\mathrm B}T$ from Monte
     Carlo and Langevin dynamics simulations. The solid line is a fit
     to the low-temperature data in order to obtain the energy barrier
     numerically. $L = 128$, $\mu_s B_z = 0.15J$, and $w = 0.032J$}
 \label{f:tau_dE_T.ps}
\end{figure}

Fig.~\ref{f:tau_dE_T.ps} shows the temperature dependence of the
reduced characteristic time for our spin chain where the dipole-dipole
interaction is not approximated by a uniaxial anisotropy as discussed
before \cite{hinzkePRB00} but taken into account using FFT methods.
Data from Monte Carlo simulations are shown as well as from Langevin
dynamics. Both methods yield identical results.  The slope of the
curve in the low temperature limit which represents the energy barrier
is 12\% lower than the predicted one. This deviation is probably due
to the local approximation of the dipole-dipole interaction underlying
Eq.\ \ref{e:e_nu}.

We will now turn to the question whether such a nucleation mode will
also be found in an extended system.  Therefore we consider 
cylindrical systems of length 128 with varying diameter $D$.
Fig.~\ref{f:ummag} shows snapshots of two simulated systems at the
characteristic time. For $D = 8$ (left side) two nuclei originate at
the sample ends leading to two domain walls which pass the system.
This is comparable to the nucleation process in a spin chain since all
spins in a plane of the cylinder are more or less parallel (also shown
in the figure) and can be describe as one effective magnetic moment
leading to the above mentioned model of Braun \cite{braunJAP99}.

The behavior changes for larger diameter above the so-called exchange
length $l_{ex}$. For a reversal mode with inhomogeneous magnetization
within the planes a domain wall has to be created. The loss of energy
due to the existence of a $180^{\circ}$ domain wall on the length
scale $l$ in a spin chain is
\begin{equation}
 \Delta E = J \sum_{i,j} (1-\cos(\theta_i-\theta_j)) \approx
 \frac{J\pi^2}{2l}
 \label{e:e_j}
\end{equation}  
under the assumption that the change of the angle $\theta$ between the
next nearest magnetic moments is constant (in a continuum description
one can also prove that this resembles the minimum energy
configuration by minimizing the Euler Lagrange equations).

\newpage
\begin{minipage}{18cm}
  \begin{figure}
  \begin{center}
   \epsfxsize=14cm
   \hspace*{1cm}
    \epsffile{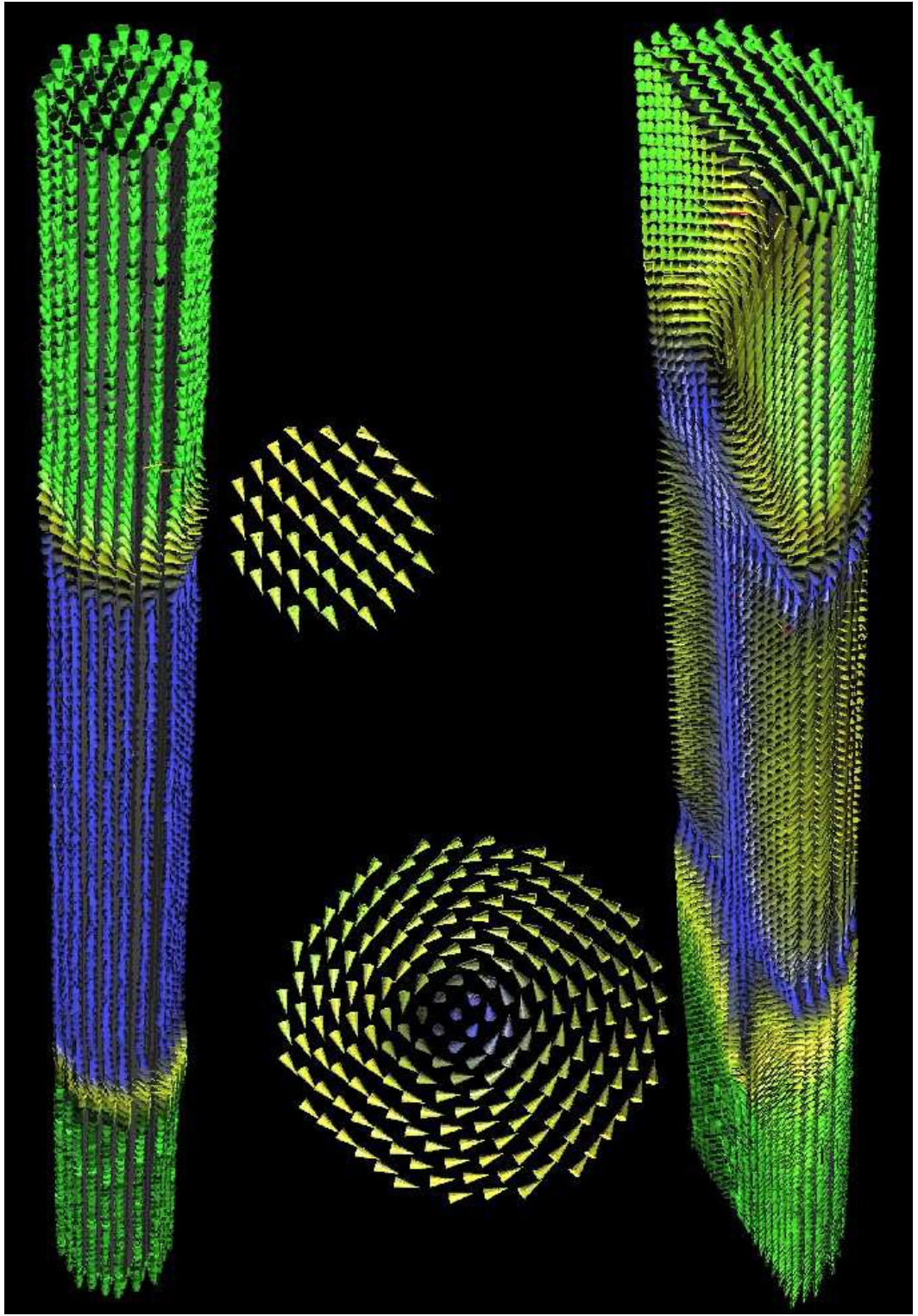}
   \end{center} 
  \caption{Snapshots of simulated spin systems at the
    characteristic time for nucleation (left) and curling (right).
    The $z$ component of the magnetic moments is also color coded
    (green corresponds to spin down, blue to spin up).  The clipping
    plane on the left side shows a layer in the upper domain wall that
    on the right side shows a layer with a magnetization vortex in the
    central curling region. $\mu_s B_z = 0.15J$, $w = 0.032J$, $L =
    128$. $D=8$ (left) and 16 (right).}
 \label{f:ummag}
\end{figure}
\vfill
\end{minipage}
\newpage

The dipolar field energy of a chain of length $l$ is at most
$3wl\zeta(3)$ where we used Riemann's Zeta function
\begin{equation}
  \zeta(3) = \sum_i^{\infty} \frac{1}{i^3}
   \label{e:e_w}
\end{equation}  
(see also \cite{huchtJMMM95} for a corresponding calculation in two
dimensions).  A comparison of Eq.~\ref{e:e_j} and Eq.~\ref{e:e_w}
yields the exchange length
\begin{equation}
  l_{\mathrm ex} = \pi \sqrt{\frac{J}{6 w \zeta(3)}}.
  \label{e:l_ex}
\end{equation}

Usually \cite{hubertBOOK98,braunJAP99} the exchange length is
calculated in the continuum limit where $3 \zeta(3) \approx 3.6$ is
replaced by $\pi$. However, we prefer the slightly deviating
expression above since its derivation is closer to the spin model
which we discuss here.  The exchange length for our material
parameters is $l_{ex} \approx 7$, so that for systems with smaller diameter
it cannot be energetically favorable for the system to build up
inhomogeneous magnetization distributions within the planes while for
wider systems another reversal process can occur namely curling
\cite{aharoniJAP96,aharoniJAP99}.

A snapshot during the reversal process in a wider particle ($D = 16$)
is also shown in Fig.~\ref{f:ummag}.  The middle part of the cylinder
is in the curling mode consisting of a magnetization vortex with a
central axis which is still magnetized up. The reversal mode shown in
this figure is still mixed up with an additional nucleation process
which started first at the sample ends.  Note, that in the clipping
plane on the right hand side one can find the exchange length as
typical length scale of the domain wall in the vortex.

\section{conclusions}     

We considered thermal activation in classical spin systems using Monte
Carlo methods with a time quantified algorithm as well as Langevin
dynamics simulations. We combined both techniques with FFT methods for
the calculation of the dipolar field. Whereas this method is
established in the context of micromagnetic equations of motion it is
less obviously applicable to Monte Carlo methods with single spin flip
dynamics. We show that it can be a good approximation for vector spins
to update dipolar fields during a Monte Carlo procedure only after
certain intervals the length of which depend on the trial step width
of the algorithm.  Since Monte Carlo methods need less computational
effort as compared to Langevin dynamics simulations its combination
with FFT methods strongly enhances the capabilities for the numerical
investigation of thermal activation.

As application we consider a model for nanowires.  We compare
numerical results for the characteristic time of the magnetization
reversal in a spin chain with the theoretical formula for the energy
barrier of soliton-antisoliton nucleation.  We found small deviations
probably due to the fact that Braun in his model approximated the
dipole-dipole interaction by a uniaxial anisotropy \cite{braunJAP99}. 
Varying the diameter in a cylindrical system we observe a crossover
from nucleation to a curling-like mode.

\acknowledgments
We thank H.\ B.\ Braun and K.\ D.\ Usadel for helpful discussions.
The work was supported by the Deutsche Forschungsgemeinschaft, and by
the EU within the framework of the COST action P3 working group 4.

\bibliographystyle{mysty}
\bibliography{/nfs/mojave/reversal/bib/Cite}
\end{multicols}
\end{document}